\documentstyle[NATO,numreferences]{crckapb}
\begin{opening}
\title{ENHANCEMENT OF KONDO TEMPERATURE \protect \\ IN NANOMETER-SIZE POINT CONTACTS}
\author{I.K. YANSON}
\author{V.V. Fisun}
\institute{B.Verkin Institute for Low Temperature Physics and Engineering\\
National Academy of Sciences, 61164 Kharkiv, Ukraine}
\author{J.A. Mydosh}
\author{J.M. van Ruitenbeek}
\institute{Kamerlingh Onnes Laboratrium, Leiden University,\\ P.O. Box 9506, 2300 RA Leiden, The Netherlands.}
\end{opening}
\runningtitle{POINT-CONTACT SPECTROSCOPY}

\begin{document}

keywords: Kondo effect, point-contact spectroscopy, size effect in
magnetic scattering

\section{Introduction}
Recently size effects in scattering of conduction electrons off 
magnetic impurities has gained a renewed interest
\cite{Berg,Gior,Chand,Zaw,Phil,YanPRL,YanFNT}. The estimate for the
characteristic size of Kondo interaction around paramagnetic impurities
embedded in a normal metal equals in order of magnitude $\xi _K\sim
v_F/T_K$ where $T_K$ is the characteristic energy scale and $v_F$ is the
Fermi velocity. In noble metals with such dissolved impurities as Mn, Cr, and
Fe $T_K$ can be quite small (down to $\sim 10^{-13}$ K in AuMn alloys \cite{Ford}) which leads to a macroscopic $\xi _K$, easily accessible in experiments. The theory predicts that at low temperatures ($T<T_K$) around each
impurity the spins of conduction electrons create a ''cloud'' which
compensates the spin of impurity \cite{Sorensen}. At higher temperatures
($T\gg T_K$) one might think that the same spatial scale determines the
logarithmic behavior of magnetic part in resistivity as a function of
temperature. It is unimportant that the average distance between
impurities, even in the most dilute alloys, is much less than $\xi _K$, since the wave functions of spin-screening conduction electrons at each impurity
are mutually orthogonal. There are a number of experiments aiming to
discover the changes of Kondo interaction across this characteristic spatial
scale \cite{Gior,Chand,Parpia}. In spite of some controversy in experimental
interpretation \cite{Chand}, a crossover distance was found \cite
{Gior,Parpia} having a negative sign for the Kondo scattering intensity as
could be naively expected. Namely, when one or more of the dimensions of a sample is decreased below a certain value, the prefactor in the Kondo logarithm
is strongly diminished. Unfortunately, this decrease cannot be connected
with the fundamental Kondo interaction, since alloys having different bulk
$T_K$ (e.g., $T_K=0.2$ K for AuFe and $T_K=30$ K for CuFe ) possess
approximately the same value for crossover scale ($\sim 10^{-5}$ cm). Around $T\sim T_K$, the same overall nonlinear temperature behavior evidences that the Kondo temperature does not depend upon the size involved \cite{Gior}.

In our point-contact measurements we have found the opposite size
effect. The samples are fabricated from Kondo alloys in the form of metallic
nanowires connecting two bulk electrodes using the technique of
mechanically controllable break junctions (MCBJ) (see below) \cite{Muller}. This
nanowire shows strong quantum oscillations of the local density of states (LDOS)
in the lateral dimensions. Since the Kondo interaction is essentially local
and quasi one-dimensional, those impurities which are located at the maxima
of LDOS manifest a noticeable increase in Kondo temperature and give
primary contribution to the resistance of a point contact.  The lower $T_K$ in
the bulk the higher is the Kondo-temperature enhancement 
\cite{YanPRL,YanFNT,YanREV,NvdPostPRB,ZarandPhys,ZarandPRB}.

\section{Experimental manifestation of Kondo scattering enhancement.}

\subsection{Comparison between phonon- and Kondo-induced 
intensities.}

The contact parameters are measured as a function of $d$ with the MCBJ
technique \cite{Muller} (see the inset in the lower panel of Fig.1).
A small rod of
the material to be investigated $(1)$ is fixed by Stycast epoxy $(3)$
to the bending beam $(4)$, which can be deflected by mechanical and piezo
drives $(5)$ leading to the pulling off of the electrodes from each other.
The notch at the center $(2$), made previously, localizes the place where
the rod is thinned in a controllable way, allowing the formation of a
nanowire.

A typical example of the first and second derivatives of the $I-V$
characteristics are plotted in Fig.1 for a CuMn (0.18 at.\%) alloy.
At zero bias the differential resistance shows the Kondo scattering maximum
which is displayed in the inset of the upper panel on an enlarged scale.
As the first characteristic parameter, we define the Kondo peak height,
$\delta R_K$. The flat minimum in the $R(V)=dV/dI(V) $ dependence, $R_0$,
is taken as the Sharvin resistance

\begin{equation}
R_0\mathrm{\, [}\Omega \mathrm{]}\approx 900/d^2 \, ,
\label{Sharvin}
\end{equation}
where $d$ [in nm] corresponds to the diameter of the narrowest cross
section of the contact and the value in the numerator is valid for copper. 

Concomitant with the Kondo peak, one can observe the
electron-phonon-interaction (EPI) spectrum $g_{PC}(eV)$ (lower panel of Fig.1)
whose shape and intensity $M$ evidences for the ballistic regime of electron
flow. As for the second characteristic parameter, we choose the intensity
of phonon-induced nonlinearity which we define as an increase of $R$ at
bias $V=20$ mV, $\delta R_{ph}$. For a copper point contact in the form
of a clean orifice \cite{Atlas} the relative phonon intensity equals

\begin{equation}
\left( \frac{\delta R_{ph}}{R_0}\right) _{V=20\mathrm{\,  mV}}=\frac
83\frac{ed}{%
\hbar v_F}\int_0^{20\mathrm{\,  mV}}g_{PC}(\omega )d\omega =4.05\times
10^{-3}d%
\mathrm{\, [nm].}  \label{PCS}
\end{equation}
For the ballistic regime of current flow, the experimental data of $%
\left( \frac{\delta R_{ph}}{R_0}\right) _{V=20\mathrm{\,  mV}}$ should
not lie below this linear dependence on contact diameter(see below)
\cite{Atlas,YanREV}.

For the weak coupling Kondo limit the theory \cite{Omel} predicts a linear
dependence of $\delta R_K/R_0$ on $d$ too, at least for the average distance
between impurities, $\langle r_0\rangle $, being much less than the contact
size $d$. Moreover, the ratio $\delta R_K/\delta R_{ph}$ should not depend
on the contact shape and regime of current flow provided the latter is
still spectroscopic (non-thermal) \cite{Duif,Atlas}.

Fig.2 shows that this is not the case in the experiment. The smaller the
size of the contact in the nanometer range, the higher is the ratio $\delta%
R_K/\delta R_{ph}$, at least for small concentrations of impurities.
One of the possible explanations might be that more and more impurities
become closer to the unitary limit of the scattering cross section, due to
the general increase of Kondo temperature for smaller $d$. Unfortunately,
the condition $\langle r_0\rangle \ll d$ is violated for smallest diameters
and another classical ''shadow'' effect may also lead to the qualitatively
the same dependence when the single impurity is situated near the contact
center \cite{Gal'perin,Kolesn}. Hence, this property alone cannot be taken
as a conclusive proof of enhanced Kondo temperature.

\subsection{Broadening of the Kondo resonance in nanometer-scale
contacts.}

Apart from the anomalous amplitude we observe a substantial increase of
width in Kondo resonance (Fig.3), for which the classical shadow-effect
cannot be responsible. This increase for contacts with sizes smaller than
approximately 10 nm is observed without any noticeable increase in width of
phonon peaks at $V\sim 20$ and $30$ mV on EPI spectrum, like that shown in
the lower panel of Fig.1.This evidences that the broadening is specific to
Kondo scattering.

For the thicker nanowires, the width of the Kondo resonance follows the
order of bulk $T_K$ for alloys CuMn and CuFe shown in Fig.3
\footnote{CuCr alloy (with bulk $T_K=2$ K) also follows this trend:
CuMn$\rightarrow $CuCr$\rightarrow $CuFe}.
Moreover, fitting the experimental Kondo resistance for CuFe by the
empirical formula $\rho _m=A-B\ln [1+(T/\theta)^2]$, where $\ln
(\theta /T_K)=-\pi [S(S+1)]^{1/2}$ \cite{Daybell}, gives the correct
$T_K\approx 30$ K for largest $d$ measured assuming the relation $eV/kT=3.63$.
This formula is valid for $T$ and $eV$ smaller than the bulk $T_K$, which is fully satisfied for CuFe alloy.

On the other hand, for the thinnest nanowires ($d\sim 2$ nm) the widths of the resonances are almost the same (enhanced $T_K^{*}\sim 100$ K) for both alloys, which can be explained by a saturation of $T_K^{*}$, approaching the electron energy subband separation $E_c=\varepsilon _F/N_c$, with $N_c$ the number of conduction channels $N_c=(k_FR/2)^2$  \cite{ZarandPC,NvdPostPRB}.

\subsection{Interplay between  Kondo temperature enhancement and Zeeman
splitting of the Kondo resonance.}

Recently Costi considered the Kondo effect in a magnetic field for strong coupling, especially in the crossover region $T\approx T_K$ \cite{Costi}. He found that a splitting of the spectral density is observed for fields larger than the critical value $H_c(T=0)\approx 0.5T_K$ in the bulk. In our experiments qualitatively the same trend is observed for increased Kondo temperature by diminishing the nanowire size $d$.

In Fig.4 we show as an example the behavior of Kondo resonance in external
(a) and internal (b) magnetic field. In the latter case the internal field
is due to spin glass ordering of a CuMn alloy with a concentration of 0.1
at.\% at $T$=0.5 K. Note that a decrease of the contact size from 13.4 to 6.4
nm (a) and 38.7 to 9.94 nm (b) fully suppress the splitting, supporting the
interpretation of an enhancement of the Kondo temperature. The suppression
of the Zeeman splitting is even more spectacular than the increase of widths
of the Kondo resonances, which is not discernible in Fig.4, due to the
broadening of the width by the magnetic field for larger diameters.

In general, if the regime of electron flow through the nanowire deviates from ballistic, then the above mentioned three manifestations of Kondo temperature enhancement disappear in the order listed in the above subsection headings. In other words, the suppression of Zeeman splitting by decreasing the contact size is the most robust property.
\pagebreak

\section{Zarand-Udvardi theory of LDOS fluctuations as a cause for the
Kondo temperature enhancement.}

Zarand and Udvardi \cite{ZarandPRB} proposed that the cause for the Kondo temperature enhancement is in the strong local electron density of states (LDOS) fluctuations inside a nanowire forming the narrowest part of a point contact.  They considered a cylindrical channel with a length $L$ and diameter $d=2R$ connecting two metallic half spaces (see inset in Fig.5a). The electron wave functions inside a cylinder are angular momentum $m$ eigenstates around the axis $z$

\[
\Psi _{\pm ,\varepsilon \lambda m}(r,z,\varphi )=e^{i\varphi
m}J_m(\lambda
r)e^{\pm ik_z(\lambda )z} \, ,
\]
where cylindrical coordinates are used. The signs $\pm $ correspond to
right- and left-going states, respectively, $\varepsilon $ denotes the
energy of the conduction electrons, and $J_m$ stands for the $m$th Bessel
function. The $z$ component of the momentum, $k_z$, can be expressed in
terms of the energy $\varepsilon $ and the radial momentum $\lambda $ of
the electron as $k_z=(2m_e\varepsilon/\hbar^2 -\lambda^2 )^{1/2}$ ($m_e$
denotes the electron mass). $\lambda $ is a continuous parameter in the
infinite half-spaces while it takes discrete values inside the tube due to
the hard-wall boundary conditions. The scattering matrices are constructed
by matching the wave functions and their derivatives at the two ends of the
cylinder \cite{ZarandPhys}.

Figure 5a shows the calculated LDOS inside the channel with $R=15$ \AA\ at $r=7$ \AA\ and $z=7$ \AA , where $z$ is measured from the left end of the point  contact. Each time a new conduction channel is opening a peak appears in the LDOS $\rho (\varepsilon ,r,z)$. Simultaneously, strong oscillations appear when the spatial coordinate is varied (Fig.5b).

Similar oscillations are the cause of the periodical pattern in a histogram of diameters obtained from large numbers of indentation-retraction cycles of MCBJ in alkali metals, due to so-called shell and supershell structure \cite{shell,super}. Although the stability of diameters for noble metal nanowires do not show the same structure at low temperature, it is quite natural to assume that strong fluctuation of LDOS exists in these metals too.

From the multiplicative renormalization group equations \cite{Fowler,ZarandPRB} the leading logarithmic terms give rise to the following expression for the ratio of the enhanced Kondo temperature $T_K^{*}$ and that referring to the bulk, $T_K$,

\begin{equation}
\frac{T_K^{*}}{T_K}=\exp \left\{ \int_{T_K}^D\frac{d\xi }{2\xi }\left[
\delta R(\xi )+\delta R(-\xi )\right] \right\}\, ,   \label{enhanced}
\end{equation}
where $\delta R(\xi )=\left[ \rho (\xi )/\rho _{bulk}(\varepsilon
_F)-1\right] $ , $\xi =\varepsilon -\varepsilon _F$ and $D$ is the bandwidth. The calculated average over the contact region $\langle T_K^{*}/T_K\rangle $ as a function of the nanowire diameter for fixed length $L=5$ \AA\ is shown in Fig.6. The increased average Kondo temperature, $T_K^{*}\sim 1-10$ K, is orders of magnitude larger than the bulk Kondo temperature for CuMn alloys, $T_K=0.01$ K, and is in reasonable agreement with the experiments (see open squares in the figure) whose Kondo temperatures were estimated by measuring the width of the Kondo resonance curves \cite{YanPRL}. A rough fitting (not shown) of the calculated data with the experimentally found power-law dependence, $T_K^{*}\sim d^{-\alpha}$, gives an exponent $\alpha =2.2\pm 0.5$, which agrees with the experimental exponent, $\alpha =2$. For comparison in Fig.6 the average $\langle T_K^{*}/T_K\rangle $ calculated by integrating the next to leading logarithmic scaling equations \cite{ZarandPRB} are shown (diamonds). Both calculations show roughly the same trend of increasing $\langle T_K^{*}/T_K\rangle $ with decreasing diameter.

In Fig.7 the calculated amplitude of the dimensionless Kondo conductance is
compared to the experimental dependence of $\delta G_K/(e^2/h)$ versus
contact diameter. Both dependences are roughly proportional to $R^2$ and
deviate strongly in amplitude and functional form ($\propto R^3$) from the
case without LDOS fluctuations (shown as a dashed line). This implies that the
Kondo resonance of the contact resistance is dominated by the magnetic
impurities having large Kondo temperatures. For the calculation, the
average $\delta G_K$ over 40 randomly chosen impurity positions were used.
Note that there is no free parameter in this calculation except for the
length of the wire, which hardly influences the results.

Large LDOS fluctuations can be generated by random scattering at the
boundary of PC as well as from the adjacent dirty banks. The following
results prove that for the observation of an enhanced Kondo temperature a
specific shape, namely a nanowire must be fabricated, as it is assumed in
the Zarand-Udvardi theory. Fig.8 shows the relative intensity of the phonon
nonlinearity, $\delta R_{ph}/R_0$, along with the Kondo peak intensity,
$\delta R_K/R_0$, for two concentration of Mn impurities in Cu.
Since the electron mean free path (including the magnetic contribution) is
larger than the contact size\footnote{The magnetic mean free path can be
estimated from the tabulated magnetic resistivity contribution in the bulk
\cite{Rizutto}.}, the regime of current flow is ballistic, which is indeed
proved by the approximate coincidence of the phonon dependences for the two
noticeably different concentrations.
Moreover, the phonon intensities are about two times larger for $d<10$ nm
than the maximal phonon nonlinearity in clean orifice-type contacts,
Eq.(\ref{PCS}), \cite{Atlas} indicated in Fig.8 by a solid straight line%
\footnote{The latter is observed as an upper limit in point-contacts produced by
''needle-anvil'' or ''edge-edge'' technique \cite{Atlas}.}.  
This means that for diameters less than 10 nm a clean nanowire, with the
length at least equal to its diameter should be created as the narrowest
part of the contact.

Considering the Kondo peak for large contact diameters, its intensity is roughly proportional to the impurity concentration evidencing that we indeed measure the scattering of conduction electrons off paramagnetic impurities. The slope of $\delta R_K/R_0(d)$ is less than for $\delta R_{ph}/R_0(d)$ for $d\leq 10$ nm in accordance to the plot shown in Fig.2 and in Fig.7.

Summarizing, since the Kondo effect is a local probe of the LDOS, the
paramagnetic impurities located close to the maxima of the LDOS inside a
nanowire give the primary contribution to the contact resistance and show
an enhanced Kondo temperature. Unfortunately, the estimates for the Kondo
temperature are still quite uncertain due to the lack of a full quantum
theory in the region where the Boltzman equation approach becomes inadequate.

\newpage\ 
\begin{figure}
\caption[]{Upper panel: Differential resistance measured for a mechanically-controlled break junction (MCBJ) for CuMn (0.18 at.\%). In the inset the magnified Kondo peak is shown. $R_0$ is a resistance connected with contact diameter $d$ via Sharvin formula (Eq.(\ref{Sharvin})). $\delta R_{ph}$ is the increase of resistance due to the phonon backscattering at $V=20$ mV. $\delta R_K$ is the Kondo-peak height. Lower panel: the second derivative of 
current-voltage characteristic of the same contact. $M$ is the maximum height of the electron-phonon interaction with a linear background subtracted. In the inset: a schematic view of the MCBJ technique. 1- the CuMn alloy; 2-notch; 3-Stycast epoxy; 4-bending beam; 5- push-pulling rod. $T=1.6$ K.}
\end{figure}

\begin{figure}
\caption[]{The ratio of the resistance increase due to Kondo and phonon scatterings as a function of the contact diameter. Triangles, and circles stand for CuMn alloys of 0.03 and 0.18 at.\%, respectively. The dotted and dashed lines serve as a guide to the eye for a continuously measured series of contacts. }
\end{figure}

\begin{figure}
\caption[]{Broadening of the Kondo-peak width with decrease of contact diameter.
The symbols and line types correspond to different contact diameters. With the
thicker lines are shown the diameters around 10 nm. Note that the width for CuMn alloy is much narrower than that for CuFe for large contact diameters in accordance with their bulk $T_K$. }
\end{figure}

\begin{figure}
\caption[]{
Suppression of the Zeeman splitting of the Kondo resonance by enhancement of the Kondo temperature with decreasing of the contact size. (a) For an external field of 9 T, the expected splitting is shown as a horizontal bar with a length 4$\mu_B$H. There is no splitting in zero field. (b) The same effect for internal magnetic field in spin glass state for an alloy with a higher concentration of impurities. $T=0.5$ K. }
\end{figure}

\begin{figure}
\caption[]{{\bf a}) Fluctuations of the LDOS inside a circular tube with $R=15$ \AA\ and $L=15$ \AA\ at the point $r=7$ \AA\ and $z=7$ \AA\ relative to the end of the tube. The dashed line shows the density of states in the bulk.  The inset schematically displays the contact geometry. {\bf b}) LDOS for the same contact with $z=7$ \AA\ and $\varepsilon =7$ eV as a function of the radius $r$. (After \cite{ZarandPRB}).}
\end{figure}

\begin{figure}
\caption[]{Calculated average relative Kondo temperature as a function of the nanowires radius $R=d/2$ of a CuMn point contact with $L=5$ \AA . Crosses and diamonds denote results obtained in the leading (Eq.\ref{enhanced})) and 
next-to-leading logarithmic orders, while open squares correspond to the experimentally observed widths in  \cite{YanPRL}. (After \cite{ZarandPRB}).}
\end{figure}

\begin{figure}
\caption[]{Size dependence of the amplitude of the dimensionless Kondo conductance. Diamonds denote the experimental data taken from \cite{YanPRL,YanFNT} while the theoretical results are indicated by crosses. The dashed line indicates the results without LDOS fluctuations $(\delta G\propto R^3)$ while the solid line corresponds to the best fit to the experimental data $(\delta G\propto R^{2.17})$ . Note that there is no free parameter in the calculation except for the length of the nanowire. (After \cite{ZarandPRB}).}
\end{figure}

\begin{figure}[tbp]
\caption[]{Relative increase of the differential resistance due to phonon (open data points) and Kondo (solid data points) scattering. Circles and triangles stand for CuMn alloys of 0.18 and 0.03 at. \%, respectively. The solid line represents the clean-orifice phonon-scattering intensity, Eq.(\ref{PCS}).
Dotted straight lines are the apparent linear fits for Kondo scattering for the same concentrations.}
\end{figure}

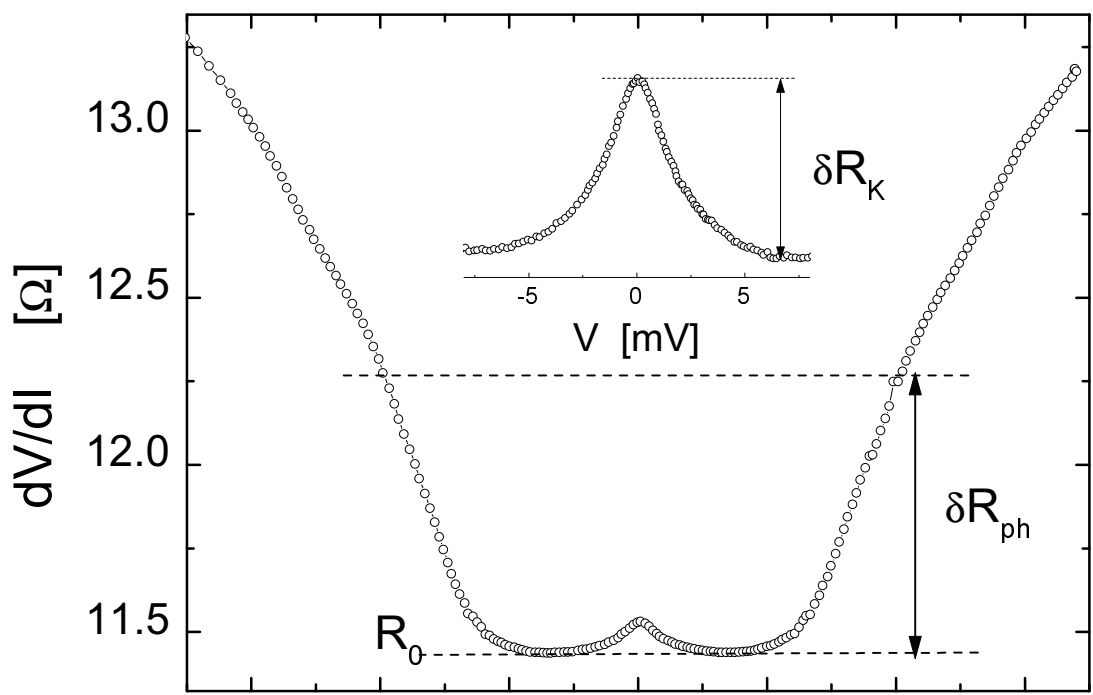
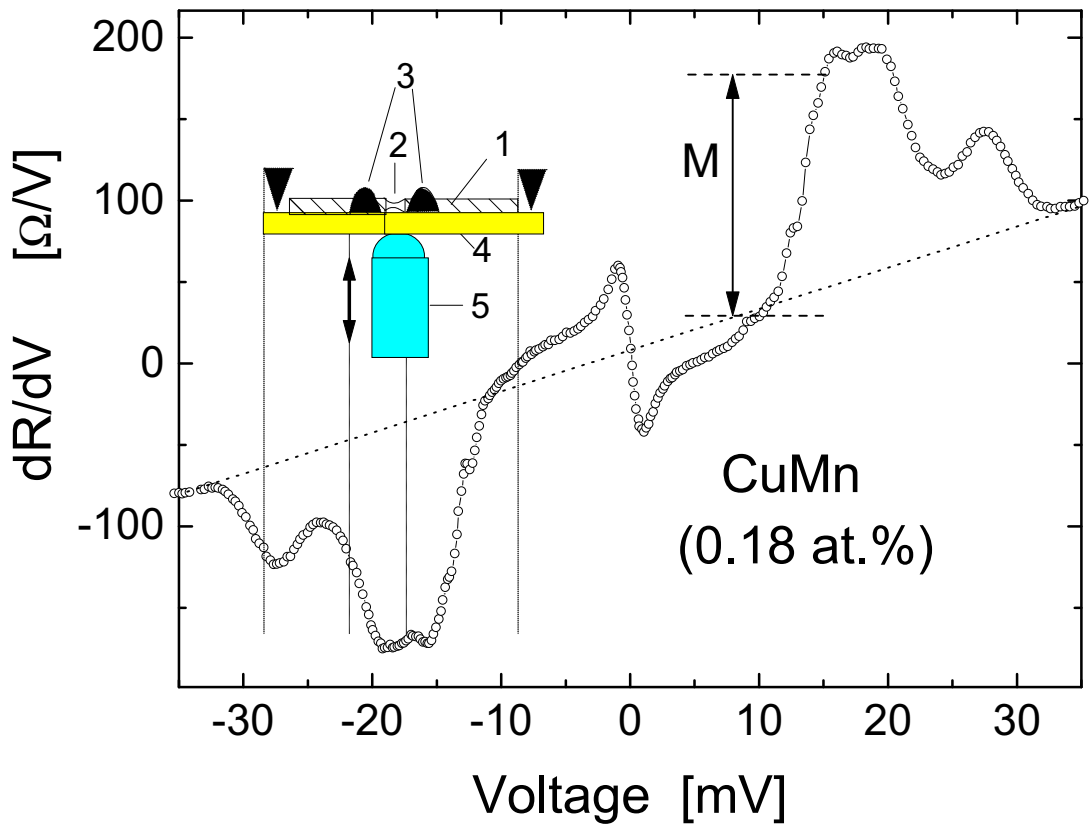
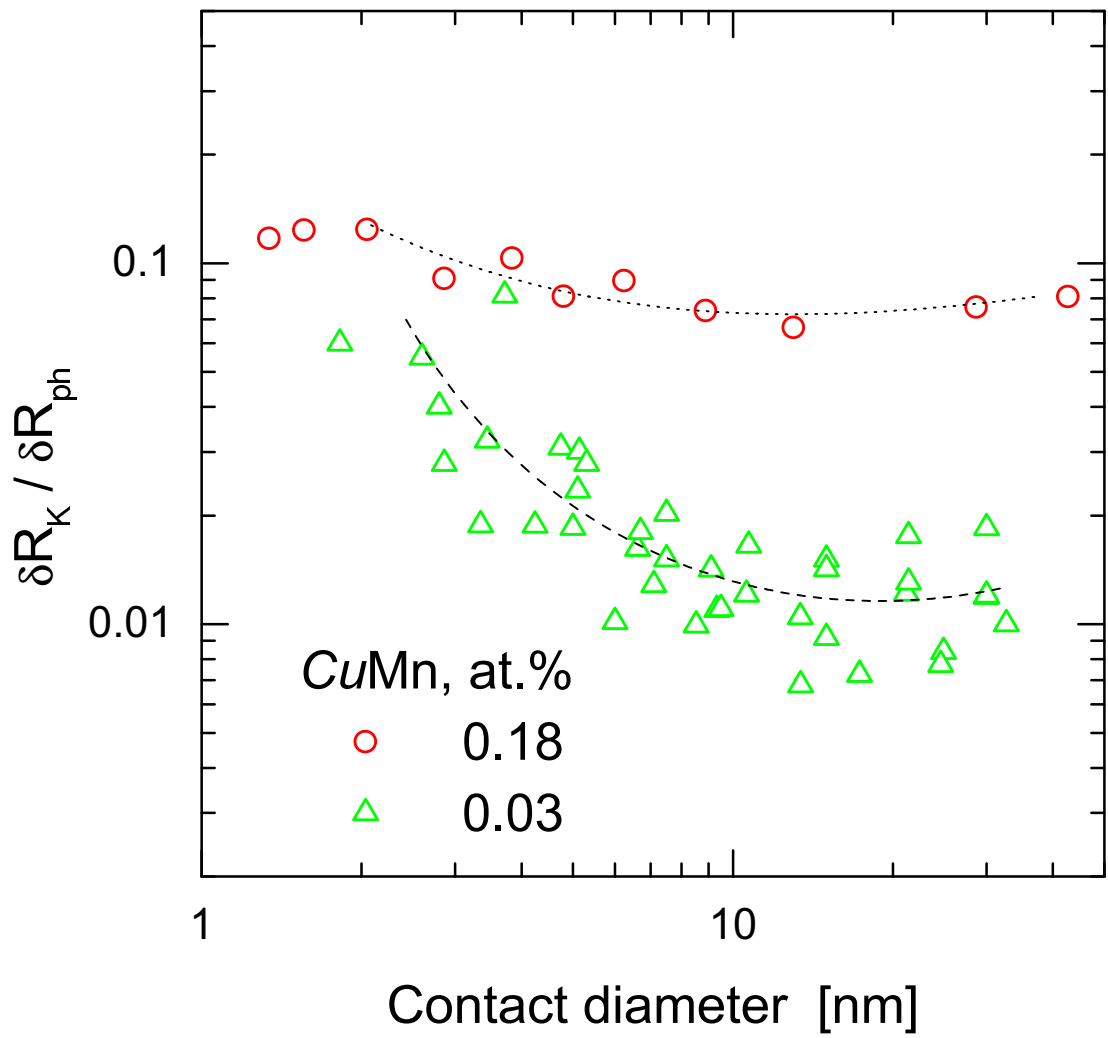
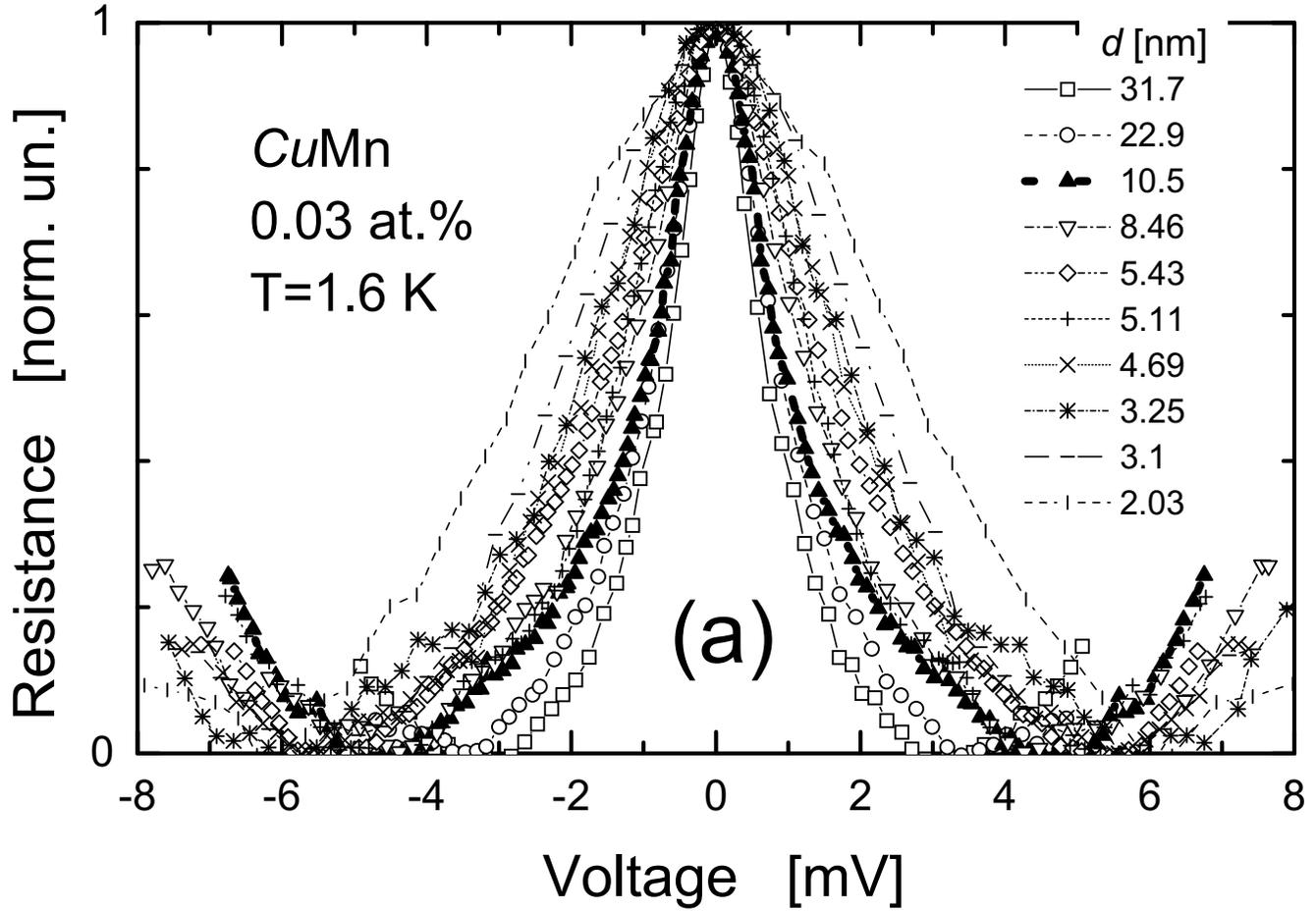
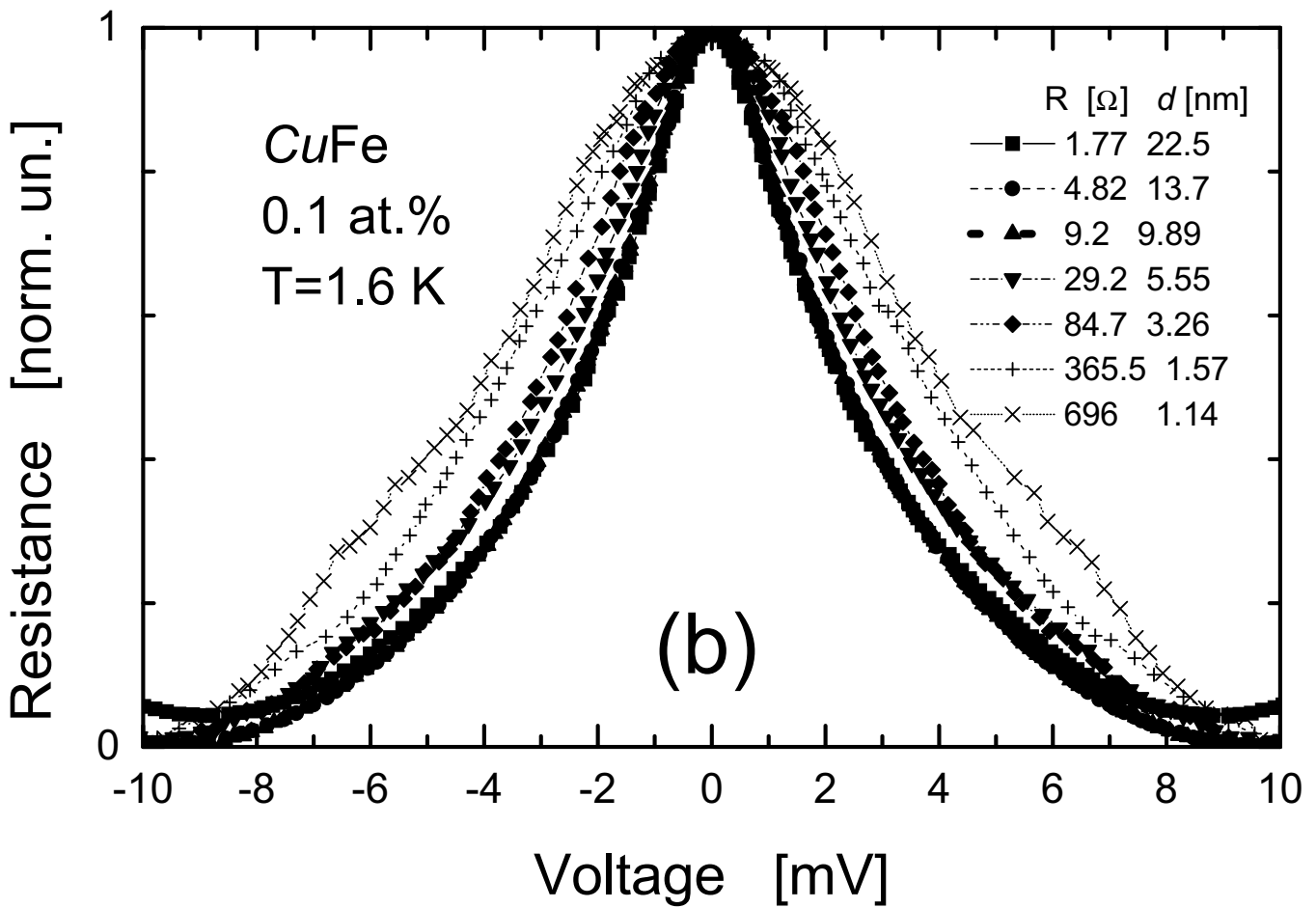
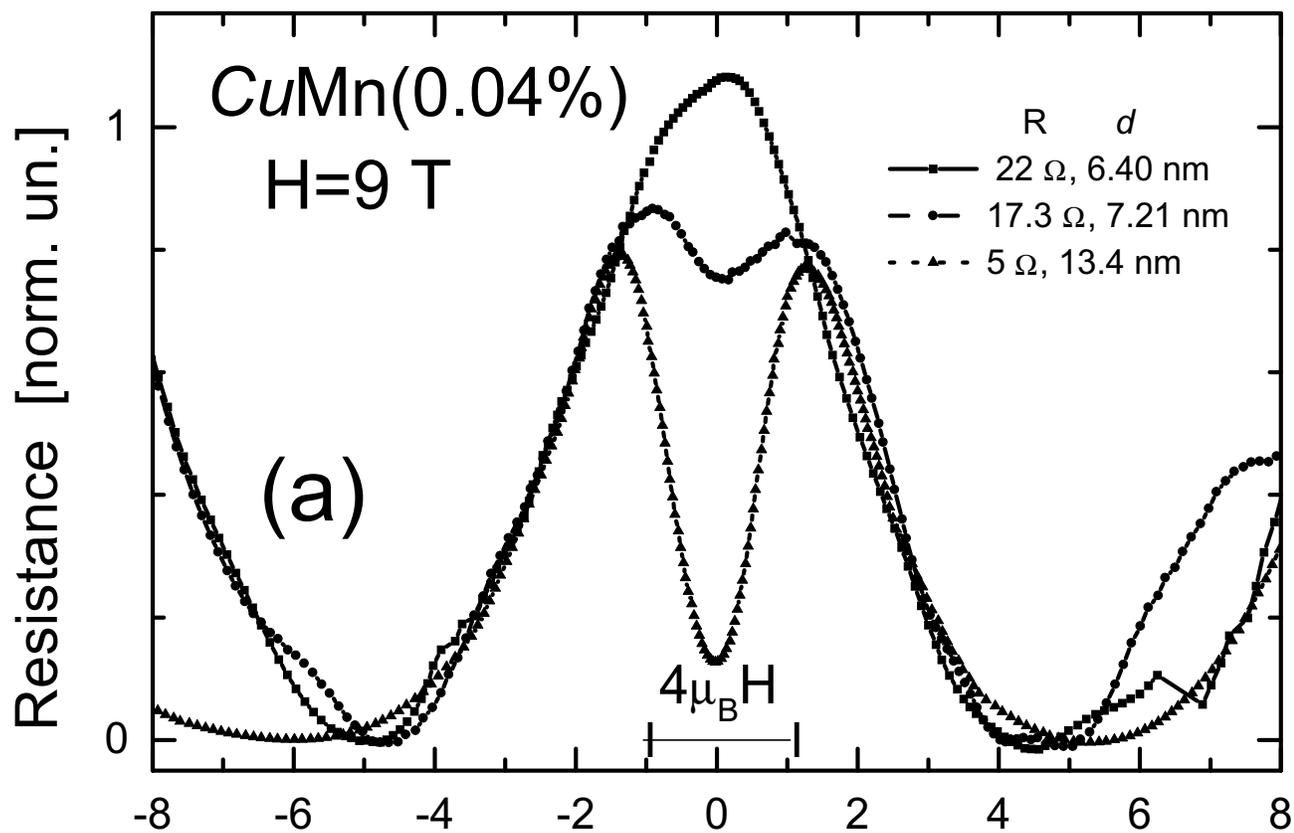
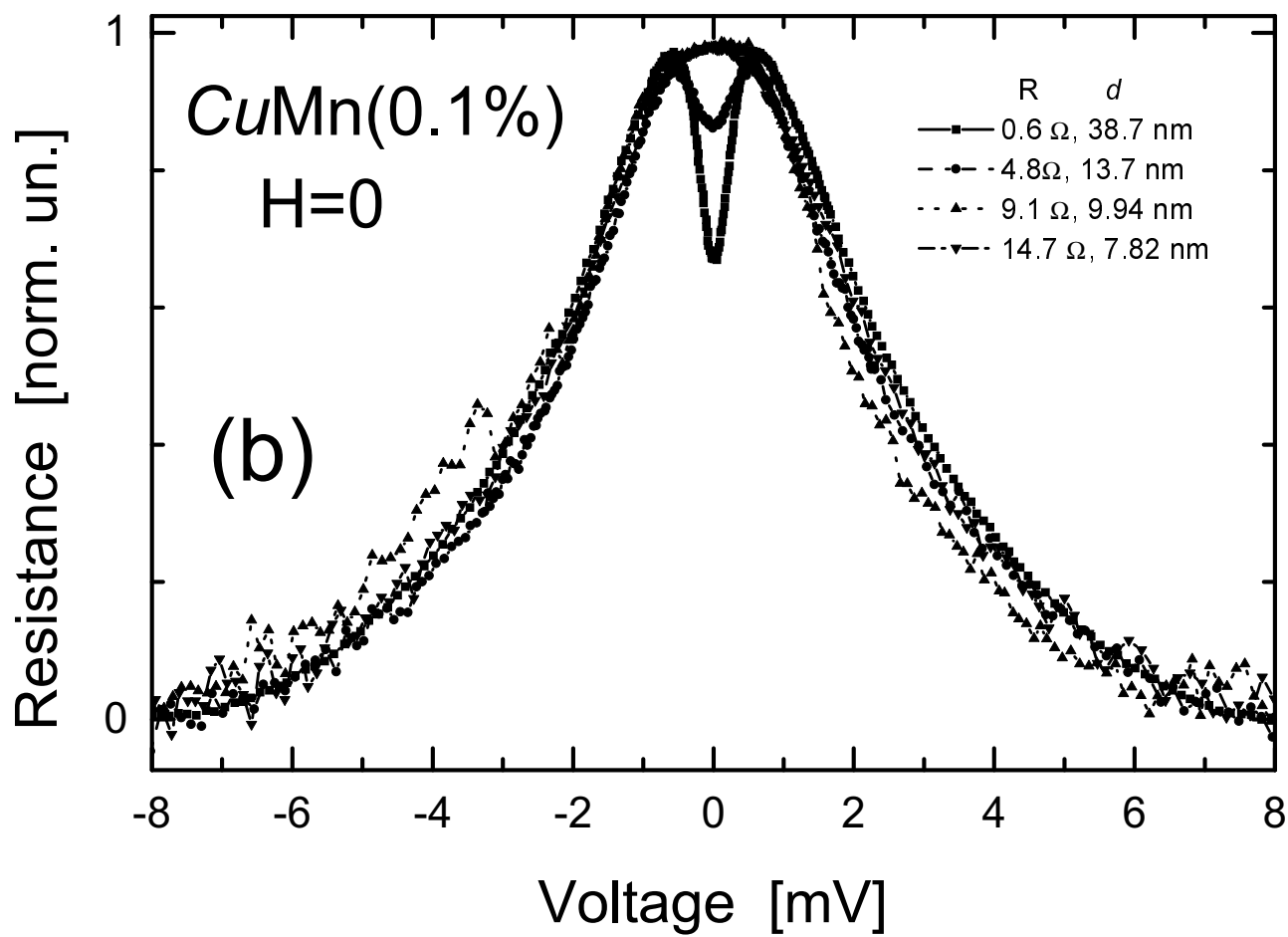
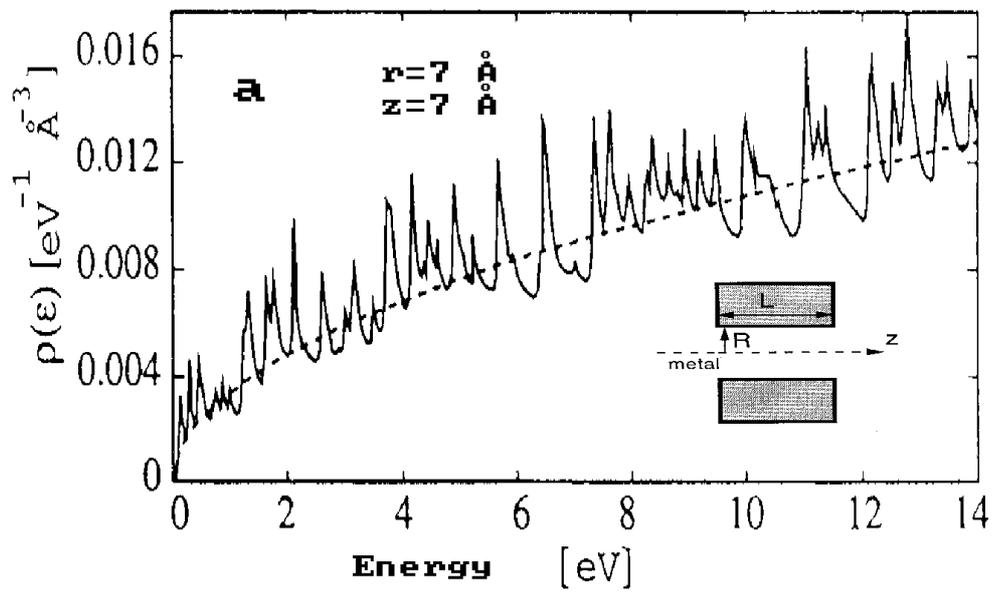
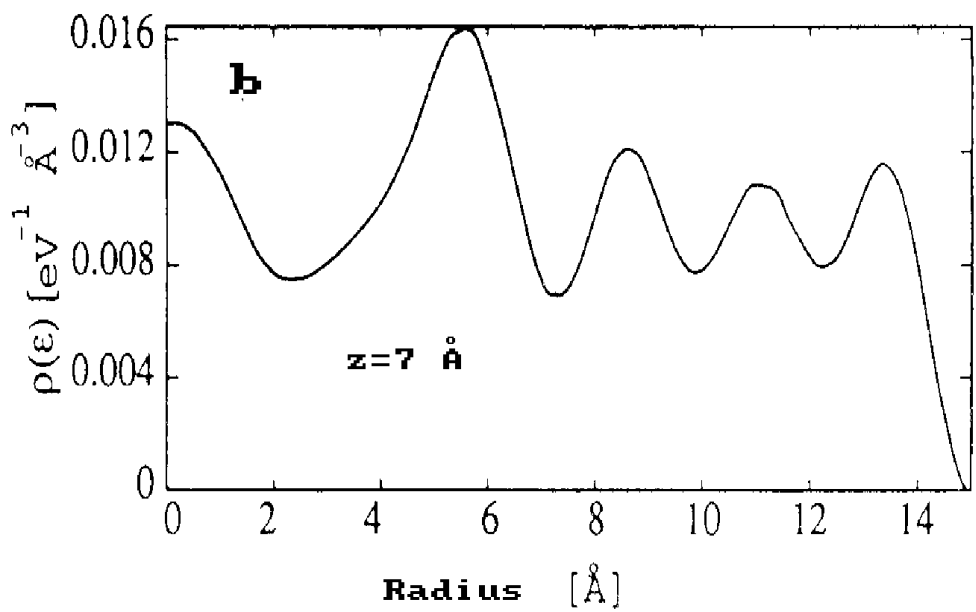
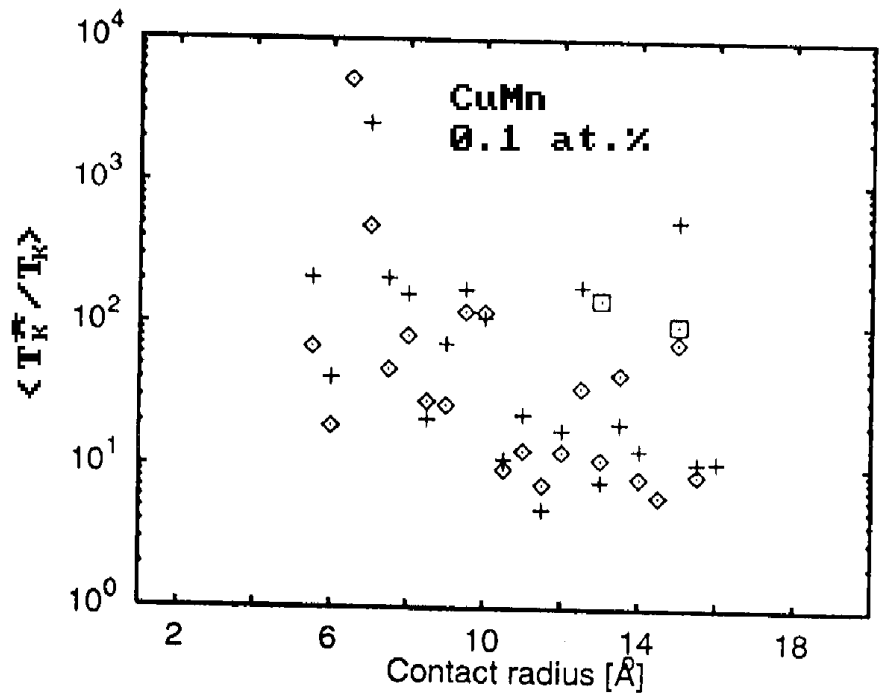
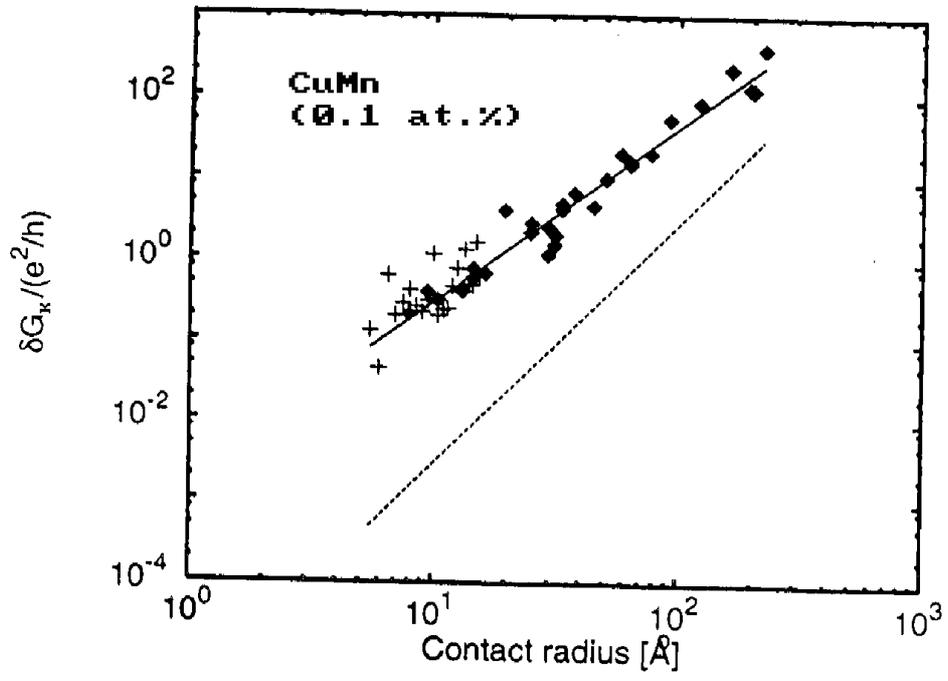
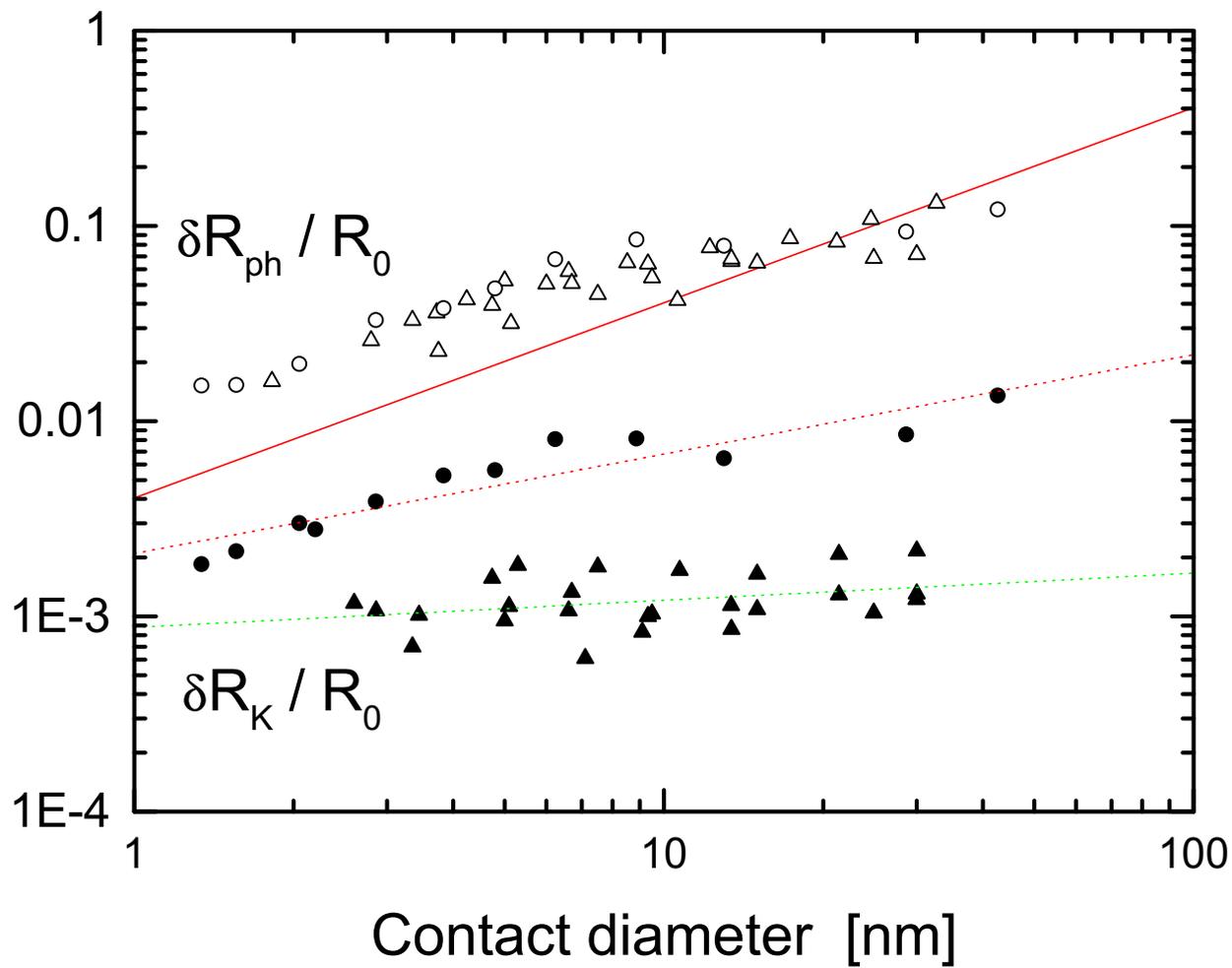
\end{document}